\documentclass[12pt]{article}
\usepackage{epsf}
\def\be{\begin{equation}}
\def\ee{\end{equation}}
\def\bea{\begin{eqnarray}}
\def\eea{\end{eqnarray}}

\def\rcn{R^{c/n}}
\def\drcn{\delta R^{c/n}}

\begin{document}
\begin{titlepage}
\begin{center}
{\Large \bf William I. Fine Theoretical Physics Institute \\
University of Minnesota \\}
\end{center}
\vspace{0.2in}
\begin{flushright}
FTPI-MINN-18/05 \\
UMN-TH-3714/18 \\
March 2018 \\
\end{flushright}
\vspace{0.3in}
\begin{center}
{\Large \bf Charged-to-neutral meson yield ratio at $\psi(3770)$ and $\Upsilon(4S)$ revisited.
\\}
\vspace{0.2in}
{\bf  M.B. Voloshin  \\ }
William I. Fine Theoretical Physics Institute, University of
Minnesota,\\ Minneapolis, MN 55455, USA \\
School of Physics and Astronomy, University of Minnesota, Minneapolis, MN 55455, USA \\ and \\
Institute of Theoretical and Experimental Physics, Moscow, 117218, Russia
\\[0.2in]

\end{center}

\vspace{0.2in}

\begin{abstract}
The recent precision measurement of the relative yield of $D^+D^-$ and $D^0 \bar D^0$ meson pairs at the $\psi(3770)$ peak in $e^+e^-$ annihilation is about 3.7 sigma away from a straightforward theoretical estimate. It is argued that combined with an even  more significant deviation of  similar ratio for the $B$ mesons at the $\Upsilon(4S)$ peak, this provides an insight into the structure of heavy mesons and requires both a form factor, corresponding to their finite `size', and a strong interaction between the mesons mesons in the isovector state $I^G(J^P) = 1^+(1^-)$. It is also argued that the data for the $D$ and $B$ mesons are possibly compatible with the heavy quark limit of the same meson-antimeson strong interaction in both systems. 
\end{abstract}
\end{titlepage}

The charged-to-neutral yield ratio in $e^+e^-$ annihilation for the $D$ meson pairs at the $\psi(3770)$ resonance and the similar ratio for the $B$ mesons at the $\Upsilon(4S)$ have attracted a considerable interest both due to the value of the ratio entering certain types of data analysis as well due to a noticed long ago~\cite{am,Lepage,be} relation to the strong-interaction structure of the flavored mesons. The recently reported result~\cite{bes18} of a precision measurement at  the $\psi(3770)$ resonance, $\rcn[\psi(3770)] = \sigma(e^+e^- \to D^+ D^-)/\sigma(e^+e^- \to D^0 \bar D^0) = (78.29 \pm 0.94)\%$,  calls for revisiting the topic of theoretical calculation of this ratio. The deviation of the ratio $\rcn$ from unity, when the mesons are produced by an isotopically neutral source (the electromagnetic current of the $c$ or $b$ quarks), is driven by the isotopic difference of the masses of heavy mesons and by the effect of the Coulomb attraction between the charged mesons. 
However the detailed contribution of these two factors depends on both the short-distance structure of the mesons as well as on the strong interaction between them in the isovector channel, i.e. with $I=1$~\cite{mv03,dlorv}. The new measured value of $\rcn[\psi(3770)]$ is in fact by about 3.7 standard deviations away from an estimate that one would obtain neglecting the structure and the interaction effects, which is also in line with a significant deviation of the world average~\cite{pdg} $\rcn[\Upsilon(4S)]=1.058 \pm 0.024$ from the result of a straightforward estimate. The purpose of the present paper is to argue that a satisfactory simultaneous description of the data on $\rcn[\psi(3770)]$ and $\rcn[\Upsilon(4S)]$ requires both the effects of the structure of the heavy mesons and of a strong interaction in the $I=1$ $P$-wave channel. Furthermore it will be argued that the data for the $D$ and $B$ mesons can be possibly related using the heavy quark limit.

The kinematical difference between the channels due to the isotopic mass splitting and the Coulomb attraction for the charged mesons, treated as point particles, gives rise to the  expression 
\be
\rcn_0 = {p^3_c \over p^3_n} \, {2 \pi \lambda \over 1 - \exp(- 2 \pi \lambda)} \, \left ( 1+ \lambda^2 \right )~,
\label{rcn0}
\ee
where $p_c$ and $p_n$ are the momenta of the charged and neutral mesons in the center of mass (c.m.) system, and $\lambda = \alpha/(2 v_c)$ is the Coulomb parameter with $v_c$ being the the velocity of one charged meson in the c.m. system. [The `extra' factor $(1+\lambda^2)$ is due to the $P$-wave production. See e.g. in the textbook \cite{ll}.]  A straightforward substitution of the known masses of the mesons 
in this formula yields (for the `zeroth approximation' ratios) $\rcn_0[\psi(3770)]= 0.7484$ and $\rcn_0[\Upsilon(4S)]=1.248$, where the c.m. energy is set to the (nominal) maximum of the corresponding resonance peak: $3773.1\,$MeV for $\psi(3770)$ and $10579.4\,$MeV for $\Upsilon(4S)$. One can see that both estimates meaningfully differ from the experimentally measured ones and these differences require an explanation. 

Clearly, if there was present a production of the mesons in both the isoscalar $I=0$ state and in the isovector one, $I=1$, the ratio would be modified by an additional factor $|A_0 - A_1|^2/|A_0+A_1|^2 \approx 1 - 4 \, {\rm Re} (A_1 / A_0)$, where $A_{0,1}$ are the corresponding isotopic amplitudes. Such effect of an `isovector contamination' is visible in precision measurements~\cite{cmd} of $K \bar K$ production at the $\phi$ resonance, where its maximal contribution to the difference of the cross sections for charged and neutral mesons is about one percent of the maximal peak cross section. This effect is due to the production of the Kaon pairs by the isovector electromagnetic current of the light $u$ and $d$ quarks and is well described by the `tail' of the $\rho$ meson~\cite{cmd}. (One can readily notice that this description is equivalent to a vector dominance model for the Kaon form factor.) At the $\psi(3770)$ mass this effect should be suppressed in the amplitude by the factor $(M^2_\phi - M^2_\rho)/(M^2_\psi - M^2_\rho) \approx 0.03$ and by an additional factor at least as small $\Lambda_{QCD}/m_c$, which accounts for the suppression of the production of the heavy charmed quarks with mass $m_c$. Thus the isovector contamination in $\psi(3770)$ is expected to be well below the current precision in $\rcn$ and is further suppressed for the heavier $B$ mesons~\footnote{The amplitude $A_1$ is non-resonant. Thus its interference with the resonant amplitude $A_0$ changes sign across the resonance, as observed~\cite{cmd} for the $\phi$ meson. Thus the isovector contamination can in principle be directly studied experimentally, although it may require a lot of data across the heavy resonance peak, e.g. the $\psi(3770)$.}.

The isotopic mass difference and the Coulomb interaction produce somewhat different effects for $D$ and $B$ mesons. For the $D$ mesons the isotopic mass difference $\Delta M = M(D^+) - M(D^0) \approx 4.78\,$MeV, combined with the fact that the $\psi(3770)$ resonance is only slightly above the threshold,  accounts for a significant suppression factor $(p_c/p_n)^3 \approx 0.6893$ while the Coulomb factor is of only mild value: 1.0890, due to a relatively small parameter $\pi \lambda \approx 0.0857$. For the $B$ mesons the relative importance of the Coulomb and mass difference effects is quite different: the Coulomb factor is  1.198, due to a larger parameter $\pi \lambda \approx 0.183$, while the small isotopic mass difference contributes only a mild factor factor 1.045.  
For a pure isoscalar source of meson pairs both factors in the expression in Eq.(\ref{rcn0}) are modified for two reasons: one being the modification of the phase relation between outgoing waves in the two channels due to the scattering phase $\delta_1$ in the isovector channel, and the other being a finite size of the mesons, which e.g. for the Coulomb interaction can be attributed to an electromagnetic form factor. 
These two effects were studied in detail~\cite{dlorv} for the $P$ wave production and in the linear order in both $\Delta M$ and $\lambda$, where both isospin violating effects are treated as a perturbation in the potential. The general formula in terms of the difference $\Delta V(r)$ of the potential for a pair of charged mesons from that for a neutral pair reads as
\be
 \rcn=  1+{ 1 \over v} \, {\rm Im}\left [ e^{2i \delta_1} \,
\int_a^\infty e^{2ipr} \, \left ( 1+ {i \over p r} \right )^2 \,
\Delta V(r) \, dr \right ]~.
\label{rcn1}
\ee
It should be noted that extending this formula to higher orders in the isospin violation is somewhat ambiguous. Indeed, the definition of the scattering phase $\delta$ is tied up to the momentum $p$ in the scattering channel, and this phase is also momentum dependent. Thus the definition of the phase $\delta_1$ in an isotopically pure state assumes that no distinction is made between $p_c$ and $p_n$~\footnote{It can be mentioned that in a similar problem in an $S$ wave this ambiguity can be somewhat relaxed in a description~\cite{lv} in terms of the scattering lengths in isotopic channels.} For a practical estimate this difficulty can be treated as follows. The only factors where the first order in the correction deviates from the full effect in Eq.(\ref{rcn0}) by more than the experimental error is in the mass effect for the $D$ mesons:
\be
1- {3 \Delta M \over v p} \approx 0.6295,
\label{fom}
\ee
compared with $(p_c/p_n)^3 \approx 0.6893$, and also the second order Coulomb effect for the $B$ mesons is slightly smaller than but still comparable with the error. Indeed,  in the first order  one finds the Coulomb factor
\be
1+\pi \lambda \approx 1.183,
\label{foc}
\ee
vs. the full result 1.198. Thus it appears sufficient to take into account the modification of the first order corrections to $\rcn$ using the formula (\ref{rcn1}) and leave the small effect of the higher orders unmodified (but still included in the calculation). 

The lower limit $a$ in the integral in Eq.(\ref{rcn1}) is a short-distance cutoff in the following sense. The difference in the potential $\Delta V$ can be defined as long as the individual channels with charged and with neutral mesons can be distinguished. However at shorter distances where the mesons overlap due to their finite spatial size, the two channels are generally strongly mixed and thus can not be distinguished. Thus any difference in the potential $\Delta V$ should vanish at short distances. This can be implemented either by introducing a form factor that takes $\Delta V$ to zero at $r \to 0$, or, as is adopted here, by introducing~\cite{dlorv} one effective cutoff distance $a$, so that $\Delta V =0$ at $r < a$. Clearly, the latter approximation corresponds to a rapid onset of the meson overlap regime at $r = a$ with the transition interval being shorter than $a$. 

The potential difference due to the mass splitting and the Coulomb interaction has the form $\Delta V = 2 \Delta M - \alpha/r$, and the integral in Eq.(\ref{rcn1}) is given as~\cite{dlorv}
\bea 
&&\int_{a}^\infty e^{2ipr} \,
\left ( 1+ {i \over p r} \right )^2 \, \, \Delta V(r) \, dr =  
\nonumber \\
&& - {\Delta M \over p} \left [ {2 \, \cos 2 \, p a \over p a}+ \sin 2 \, p a +
i \, \left ( {2 \sin 2 \, p a \over p a} - \cos 2 \, p a \right) \right ]  \nonumber \\
&& +  \alpha \left [ \left ({\cos 2 pa \over 2
(pa)^2}+{\sin 2 pa \over pa}-{\rm Ci}(2pa) \right ) + i \, \left ( {\pi \over
2}-{\cos 2 pa
\over pa} + {\sin 2 pa \over 2 (pa)^2} - {\rm Si}(2pa) \right ) \right ]~, 
\label{intv}
\eea
where Si$(z)$ and Ci$(z)$ are the integral sine and cosine functions. The real part of this expression has a singularity at $p \to 0$. The physically meaningful expression in Eq.(\ref{rcn1}) is however finite in this limit, since the singular real part of (\ref{intv}) is multiplied by $\sin 2 \delta_1$, and at small $p$ the $P$-wave scattering phase vanishes as $p^3$. (See e.g. in Ref.~\cite{ll}.)

A comparison with the data is performed here in terms of the deviation $\drcn = \rcn - \rcn_0$ of the measured $\rcn$ from that given by the formula (\ref{rcn0}):  $\drcn[\psi(3770)]= 0.0345 \pm 0.0094$ and 
$\drcn[\Upsilon(4S)]= - 0.190 \pm 0.024$. As discussed, this deviation is attributed here to the effects of strong interaction scattering phase $\delta_1$ in the isovector state and of the short distance cutoff $a$in the linear in the mass splitting and the Coulomb interaction terms in the correction [Eq.(\ref{rcn1})]. Accordingly, the working expression for $\drcn$ reads as
\bea
&&\drcn = {3 \Delta M \over p v} - {\pi \alpha \over 2 v_c} +  \nonumber \\
&&{1 \over v} \,  {\rm Im} e^{2i \delta_1} \,\left \{ - {\Delta M \over p} \left [ {2 \, \cos 2 \, p a \over p a}+ \sin 2 \, p a +
i \, \left ( {2 \sin 2 \, p a \over p a} - \cos 2 \, p a \right) \right ] \right . \nonumber \\
&&+ \left. \alpha \left [ \left ({\cos 2 pa \over 2
(pa)^2}+{\sin 2 pa \over pa}-{\rm Ci}(2pa) \right ) + i \, \left ( {\pi \over
2}-{\cos 2 pa
\over pa} + {\sin 2 pa \over 2 (pa)^2} - {\rm Si}(2pa) \right ) \right ] \right \} ~, 
\label{wf}
\eea
where the first two terms on the r.h.s correspond to the subtraction of the unmodified first order in $\Delta M$ and $\alpha$ contribution that is included in $\rcn_0$ in Eq.(\ref{rcn0}). It can be noticed that in this expression $p$ and $v$ are the momentum and velocity averaged over the charged and neutral mesons: $p \approx 270\,$MeV, $v \approx 0.143$ for the $D$ mesons at $\psi(3770)$ and  $p \approx 329\,$MeV, $v \approx 0.0621$ for the $B$ mesons at $\Upsilon(4S)$.

\begin{figure}[ht]
\begin{center}
 \leavevmode
    \epsfxsize=8cm
    \epsfbox{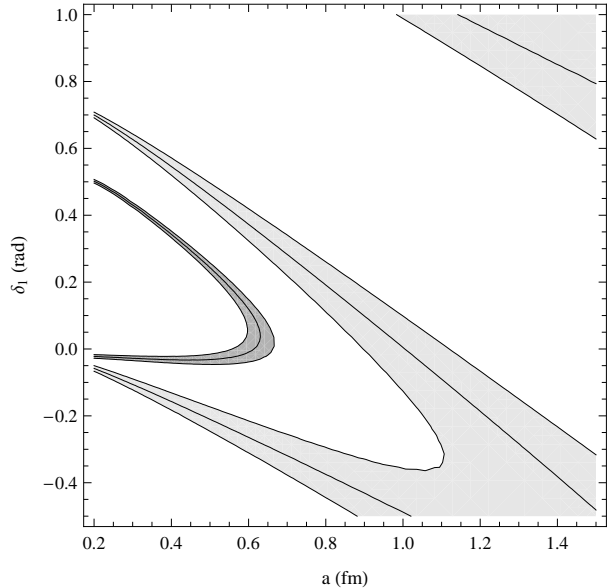}
    \caption{The one sigma contours for the fit to the data on $\rcn[\psi(3770)]$ (the darker shading) and $\Upsilon(4S)$ (the lighter shading) using Eq.(\ref{wf}).}
\end{center}
\end{figure}

The fit of the expression in Eq.(\ref{wf}) to the data in terms of the scattering phase $\delta_1$ and the distance parameter $a$ is shown in Fig.~1. It is instructive to compare the implications from the $\psi(3770)$ and $\Upsilon(4S)$ data in terms of the structure and interaction of the heavy mesons.
For such comparison one can recall that the form factor scale (the size of the meson) and the interaction potential between the mesons are constant in the heavy quark limit. Thus the best we can do at the present level of accuracy is to rely on this limit for both the $b$ and $c$ quarks and to neglect any differences in effective size or the meson-antimeson interaction between the $D$ and $B$ mesons. In particular this implies that the parameter $a$ in our formulas is the same for both systems. Unlike the interaction potential, the scattering phase does not go to a constant in the heavy quark limit, and depends on the (heavy) mass and the momentum of the mesons, and naturally increases (in absolute value) with the mass and the momentum. Thus one can expect that the value of $|\delta_1|$ for the $B$ mesons at the $\Upsilon(4S)$ resonance energy is significantly larger than that for the $D$ mesons at the $\psi(3770)$ peak. The contours in Fig.~1 do not contradict this behavior: at a constant $a < 0.6 - 0.7\,$fm the fits are compatible if $|\delta_1|$ for $B$ mesons is significantly larger than for the charmed mesons. Clearly, this agreement is possible only if the scattering phase is non vanishing, so that both the effects of a finite $a$ and of the interaction are required in order to explain the ratio $\rcn$ at both $\psi(3770)$ and $\Upsilon(4S)$. 

\begin{figure}[ht]
\begin{center}
 \leavevmode
    \epsfxsize=8cm
    \epsfbox{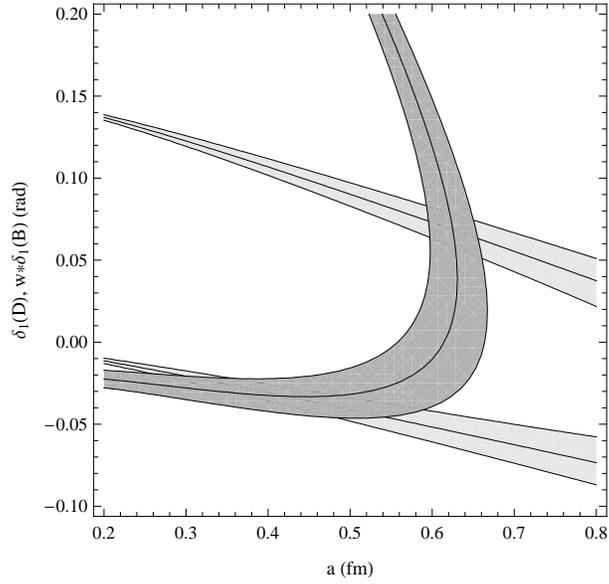}
    \caption{The one sigma contours for the fit to the data on $\rcn[\psi(3770)]$ (the darker shading) and $\Upsilon(4S)$ (the lighter shading) using Eq.(\ref{wf}). The scattering phase for the $B$ mesons is scaled by the factor $w$ in Eq.(\ref{dbsw}).}
\end{center}
\end{figure} 

One can attempt to quantify the comparison of the scattering phases between the $B$ mesons and the $D$ mesons by treating these phases as small so that they can be evaluated in the Born approximation, i.e. in the first order in the interaction potential (in the I=1 state) $V_1(r)$. The scattering phase in the wave with angular momentum $\ell$ is then found from the formula~\cite{ll}
\be
\delta^{(\ell)} = - {\pi M \over 2} \int_0^\infty V_1(r) \left [ J_{\ell+1/2} (p r) \right ]^2 \, r \, dr~,
\label{dl}
\ee
where $J_\nu(z)$ is the standard Bessel function. If the effective range of the potential $V_1$ can be considered small as compared to $1/p$, one readily estimates the scaling behavior of the $P$-wave phase:
\be
\delta_1 \propto M \, p^3.
\label{dsc}
\ee
One can thus attempt to directly superimpose the $(a, \delta_1)$ contours for the two fits by rescaling the scattering phase for the $B$ mesons by the factor 
\be
w = (M_D/M_B)\,(p_D/p_B)^3 \approx 0.2~.
\label{dbsw}
\ee
The result of such comparison using a rescaled scattering phase is shown in Fig.~2. One can certainly notice that there is a compatibility region at a value $a \approx 0.6$ with a positive scattering phase, i.e. the potential $V_1$ corresponding to attraction, as well as another overlap region at a shorter scale $a$ and negative $\delta_1$ corresponding to a repulsion between the mesons. Although the solution with a longer parameter $a$ appears to be more in line with the perceived size of the mesons, given the current uncertainty, the solution with negative scattering phase can not be discarded.

In summary. The experimentally observed charge-to-neutral yield ratio $\rcn$ for the $D$ mesons at the $\psi(3770)$ resonance and for the $B$ mesons at $\Upsilon(4S)$ definitely deviates from the formula (\ref{rcn0}) for point-like non-interacting particles. The sign and the amount of deviation, different for the two systems, can be explained by the modification~\cite{dlorv} of the corrections to $\rcn$ by the finite size of the mesons and their scattering in the isovector state. It is argued here that both types of modification are necessary and that the (seemingly very different) results for $\rcn$ at the   $\psi(3770)$ and $\Upsilon(4S)$ resonances can be made compatible using a heavy quark limit scaling from $D$ to $B$ mesons of the parameters of interaction between the mesons.

This work is supported in part by U.S. Department of Energy Grant No.\ DE-SC0011842.

 \end{document}